# Superconductivity protected by spin-valley locking in ion-gated MoS$_2$

Yu Saito[1]*, Yasuharu Nakamura[2, 3], Mohammad Saeed Bahramy[1,4], Yoshimitsu Kohama[5],

Jianting Ye[1,4], Yuichi Kasahara[3], Yuji Nakagawa[1], Masaru Onga[1], Masashi Tokunaga[5],

Tsutomu Nojima[6], Youichi Yanase[3,7] and Yoshihiro Iwasa[1,4]*

[1] *Quantum-Phase Electronics Center (QPEC) and Department of Applied Physics,*

*The University of Tokyo, Tokyo 113-8656, Japan*

[2] *Graduate School of Science and Technology, Niigata University, Niigata 950-2181, Japan*

[3] *Department of Physics, Kyoto University, Kyoto 606-8502, Japan*

[4] *RIKEN Center for Emergent Matter Science (CEMS), Wako 351-0198, Japan*

[5] *International MegaGauss Science Laboratory, Institute for Solid State Physics,*

*The University of Tokyo, Kashiwa 277-8581, Japan.*

[6] *Institute for Materials Research, Tohoku University, Sendai 980-8577, Japan*

[7] *Department of Physics, Niigata University, Niigata 950-2181, Japan*

*Corresponding author: saito@mp.t.u-tokyo.ac.jp (Y.S.), iwasa@ap.t.u-tokyo.ac.jp (Y.I.)

Symmetry-breaking has been known to play a key role in noncentrosymmetric superconductors with strong spin-orbit-interaction (SOI) [1-6]. The studies, however, have been so far mainly focused on a particular type of SOI, known as Rashba SOI[7], whereby the electron spin is locked to its momentum at a right-angle, thereby leading to an in-planar helical spin texture. Here we discuss electric-field-induced superconductivity in molybdenum disulphide ($MoS_2$), which exhibits a fundamentally different type of intrinsic SOI manifested by an out-of-plane Zeeman-type spin polarization of energy valleys[8-10]. We find an upper critical field of approximately 52 T at 1.5 K, which indicates an enhancement of the Pauli limit by a factor of four as compared to that in centrosymmetric conventional superconductors. Using realistic tight-binding calculations, we reveal that this unusual behaviour is due to an inter-valley pairing that is symmetrically protected by Zeeman-type spin-valley locking against external magnetic fields. Our study sheds a new light on the interplay of inversion asymmetry with SOI in confined geometries, and its unprecedented role in superconductivity.

$MoS_2$ is a member of the layered semiconducting transition metal dichalcogenides (TMDs) [11], which have been attracting widespread attention as two-dimensional (2D) materials beyond graphene, owing to their multiple functionalities with potential applications such as atomically thin electronics[12-14], photonics[15] and valleytronics devices utilizing a coupled spin and valley degree of freedom[16-18]. Also, $MoS_2$ is becoming a new platform for investigating quantum physics for example with quantum oscillation[19] and electric-field-induced superconductivity[20]. The unit cell of $MoS_2$ is composed of two formula units, in each of which one Mo atom is sandwiched between two S atoms, forming an S–Mo–S monolayer stacking along the $c$-axis with $D_{3h}$ symmetry (Fig. 1a). In the isolated monolayer, in-plane inversion symmetry is broken, causing out-of-plane spin polarization together with effective valley Zeeman fields, namely, valley-dependent Zeeman-type spin polarization at zero magnetic field [8-10, 21]. This Zeeman-type spin splitting reaches 3 meV[22] and 148 meV[8] at the bottom of conduction band and the top of the valence band, respectively, both of which are located at the K-points, the corner of the hexagonal first Brillouin zone shown in Fig. 1b. Such a zero-field spin splitting is not observed in bulk $MoS_2$ with $D^4_{6h}$ symmetry[23, 24]. Also, this spin splitting changes its sign at the –K point, because the K and –K points are connected by the time-reversal operation. Such a spin splitting unique to monolayer $MoS_2$ originates from the fairly strong SOI of transition metal $d$-orbitals, and is commonly observed in the group VI of TMD semiconductors[8, 9]. This valley-dependent spin polarization is in marked contrast to the in-plane momentum-dependent spin polarization caused by the Rashba-type SOI[7].

A noncentrosymmetric system with considerable SOIs is an ideal platform for exotic superconductivity, and in fact, superconductivity occurring in the Rashba-type band structure has been intensively investigated on a variety of systems[1-4], together with the effect of spin-momentum locking. However, the Zeeman-type spin polarization on superconductivity has never been discussed so far. Here, we investigate electric-field-induced superconductivity in $MoS_2$ by using an electric-double-layer transistor (EDLT) configuration (Fig. 1c), which creates a high density two-dimensional electron system (2DES) on the surface (Fig. 1d) without introducing extrinsic disorder, thereby offering novel opportunities to search for new types of exotic superconductivity[25-27].

In order to extract anomalous features of electric-field-induced superconductivity at the highly crystalline multilayer $MoS_2$ surface, we fabricated an EDLT structure with a 20-nm-thick flake, and then performed magnetotransport measurements. The $MoS_2$-EDLT underwent a superconducting transition at a gate voltage of $V_G = 6.5$ V and a sheet carrier density of $n_{2D} = 1.5 \times 10^{14}$ cm$^{-2}$ at 15 K (Fig. 2a). A critical temperature, $T_c$, of this device was 9.7 K, as defined at the mid-point of the transition with $R_{sheet}$ being 50% of the normal state sheet resistance at 15 K. This carrier density is slightly larger than the optimum value in the dome-shaped phase diagram[20, 28, 29]. The electrochemical reaction is unlikely even at high gate voltages up to 6.5 V (see Supplementary Section I), according to the reversibility and the absence of hysteresis in the transfer curve (Supplementary Fig. 1). Figure 2, b and c, display a zoom-up of the resistive transition in the low temperature region under the application of perpendicular and parallel magnetic fields from 0 to 9 T , respectively. The superconducting state is completely quenched at 9 T for the perpendicular magnetic fields (Fig. 2b), whereas it remains almost unchanged in the in-plane magnetic field geometry (Fig. 2c). This behaviour means a substantially large anisotropy in the superconductivity. Figure 2d shows the angular dependence of the upper critical field, $H_{c2}(\theta)$, at 9.6 K ($\theta$ represents the angle between the $c$-axis of the crystal and applied magnetic field directions). Here, $H_{c2}(\theta)$ is determined by the mid-point of the resistive transition, as well. A cusp-like peak in the inset of Fig. 2d is well described by the 2D Tinkham model[30] (and not by the 3D anisotropic mass model) as frequently observed in interfacial superconductivity[31-33]. In addition, the temperature, $T$, dependence of $H_{c2}$ for both the out-of-plane and in-plane magnetic fields (Fig. 2e) are well fitted by the phenomenological 2D Ginzburg–Landau (GL) model,

$$\mu_0 H_{c2}^{\perp} = \frac{\Phi_0}{2\pi\xi_{GL}(0)^2}\left(1 - T/T_c\right) \text{ and } \mu_0 H_{c2}^{\parallel} = \frac{\Phi_0\sqrt{12}}{2\pi\xi_{GL}(0)d_{SC}}\sqrt{1 - T/T_c} ,$$ where $\Phi_0$, $\xi_{GL}(0)$, and $d_{SC}$

denote a flux quantum, the in-plane GL coherence length at $T$ = 0 K, and the effective thickness of superconductivity, respectively. We find $\xi_{GL}(0)$ = 8.0 nm and $d_{SC}$ = 1.5 nm. It is noted that the extremely sharp rise of $H_{c2}^{\parallel}(T)$ near $T_c$ shows a marked contrast to that in the conventional bulk layered superconductors such as Cs-doped $MoS_2$[34], demonstrating that the present system is extremely 2D in nature. In fact, $H_{c2}^{\parallel}(T)$ can seemingly go far beyond the Pauli limit, $H_P^{BCS}$, for weak coupling Bardeen–Cooper–Schrieffer (BCS) superconductors, $H_P^{\mathrm{BCS}} = \Delta_0 / \sqrt{2} k_B T_c = 1.86 T_c = 18$ T, where $k_B$ and $\Delta_0$ are the Boltzmann constant and the superconducting gap at $T = 0$ K, based on the BCS theory, respectively.

To investigate $H_{c2}^{\parallel}$ at much lower temperatures, we measured the magnetoresistance of another MoS$_2$-EDLT applying pulsed magnetic fields up to 55 T (see Supplementary Section II and Supplementary Fig. 2). A clear resistance drop at a $T_c$ of 6.5 K was observed, which was defined by the temperature where $R_{\mathrm{sheet}}$ reached 75% of the normal state sheet resistance, indicating a superconducting signature, although the MoS$_2$-EDLT used for the high magnetic field measurements ($n_{2D} = 8.5 \times 10^{13}$ cm$^{-2}$ at $V_G = 5.5$ V and $T = 15$ K) did not exhibit zero resistance. Figure 3, a and b, display the magnetoresistance of the MoS$_2$-EDLT for out-of-plane and in-plane magnetic fields, respectively, at several temperatures between 1.5 and 8.0 K. In the out-of-plane magnetic field geometry, the superconducting state is completely destroyed by the application of more than 5 T magnetic fields. On the other hand, for the in-plane magnetic fields, the superconductivity is not completely suppressed nor does it revert to the normal state even by applying a 55 T magnetic field at 1.5 K. We summarize both $H_{c2}^{\parallel}(T)$ and $H_{c2}^{\perp}(T)$ in Fig. 3c. We note that $H_{c2}^{\parallel}(T)$ increases with decreasing temperature and eventually saturates approximately 52 T at 1.5 K, which is more than four times larger than $H_P^{\mathrm{BCS}} = 12$ T. Because the orbital limit is supposed to be large owing to the confined geometry by the EDLT, the saturating behaviour of $H_{c2}^{\parallel}(T)$ at low temperatures is suggestive of the Pauli limit, as seen in the Pauli limited superconductor[35].

The enhancement of $H_{c2}$ in a dirty-limit superconductor with strong SOI has been discussed in terms of the spin-orbit scattering caused by disorder. This is expected to cause randomization of electron spins, and thus result in suppression of the effect of spin paramagnetism[36-38]. To evaluate the contribution of this effect, we fitted our $H_{c2}^{\parallel}(T)$ data by using the microscopic Klemm–Luther–Beasely (KLB) theory[38], which is applicable to dirty-limit layered superconductors with strong SOI ($l << \xi_{Pippard}$ and $\tau << \tau_{SO}$, where $l$, $\xi_{Pippard}$, $\tau$ and $\tau_{SO}$ is the mean free length, the Pippard coherence length, the total scattering time and the spin orbit scattering time, respectively). Our $H_{c2}^{\parallel}(T)$ data are well fitted by the KLB theory (Supplementary Fig. 3); however, we found that in any cases the value of $\tau$ is larger than that of $\tau_{SO}$ ($\tau > \tau_{SO}$) (Supplementary Table 1). This is an unphysical situation which contradicts with the initial assumption required for this theory ($\tau << \tau_{SO}$). Thus, the model with the effect of spin orbit scattering does not explain for the enhancement of $H_{c2}^{\parallel}$, consistently.

To find a more plausible origin of the enhancement of the Pauli limit in the present system, we performed a set of *ab-intio*-based tight-binding supercell calculations on bulk $MoS_2$, incorporating the near surface band bending effect via an electrostatic potential term obtained by self-consistently solving the Poisson equation (details in Supplementary Section IV). Our calculations suggest that under the application of a strong electric field, a high density 2DES is created at the surface of $MoS_2$. As schematically shown in Fig. 1d, this results in formation of an accumulation layer, which is effectively confined within the topmost $MoS_2$ layer[39, 40], indicating that gate-induced noncentrosymmetric quasi-single-layer superconductivity is realized in our system. Such a "quasi-single-layer" 2DES, therefore, ought to have an effective $D_{3h}$ symmetry, leading to many interesting features in the momentum space. For example, once a positive gate voltage is switched on, the conduction band minimum shifts to the $\pm$K points[40]. This situation is in stark contrast to the case of bulk $MoS_2$, where the conduction band minima are located at six symmetrically equivalent $k$ points along $\Gamma$-K directions, also known as the T (or Q) points[23, 24]. Accordingly, the electric-field-induced 2D superconductivity in $MoS_2$ is expected to be solely mediated by the $\pm$K valleys, and thus the most likely ground state of the Cooper pair should be the inter-valley pairing between the +K and –K valleys in order to maintain the zero-momentum for the center-of-mass of the Cooper pairs. Note that the intra-valley spin-singlet Cooper pairs are not stabilized in the presence of the Zeeman-type SOI, which requires the non-zero momentum.

At a sheet carrier density of $n_{2D} = 8.7 \times 10^{13}$ cm$^{-2}$, which is nearly the same as the value in the high-field measurement, the bands are spin-split by ~ 3 meV at the $\pm$K points, at zero magnetic field. Slightly away from the K point, these spin-split bands cross each other such that the splitting becomes ~ 13 meV at the Fermi level. The corresponding band dispersion and spin texture at the Fermi surface are shown in Fig. 4, a and b, respectively. All these features of the band structure are qualitatively equivalent to those in the monolayer MoS$_2$ derived from the tight-binding method[41] and the $\boldsymbol{k \cdot p}$ model[42]. This agreement indicates that bulk or multilayer TMDs under a strong electric field can effectively behave as monolayers. Such a monolayer-like behaviour has been already experimentally demonstrated in bilayer systems, exhibiting circularly polarized photoluminescence under an electric field[43], and in bulk systems showing gate-induced weak anti-localization behaviour in magnetoconductance[21]. In addition to these works, the recent optical measurement on WSe$_2$ multilayers have shown that these systems can emit an electrically switchable circularly polarized electroluminescence[16]. The circularly polarized luminescence is believed to be a unique feature of the monolayer. Hence, the observation of the same phenomenon in a gated multilayer system provides strong evidence that the transition metal dichalcogenides such as MoS$_2$ can behave like a monolayer under a electric field.

(改行)As shown in Fig. 4a, each band is almost fully out-of-plane spin polarized. It is worth noticing that the in-plane Rashba-type component, which originates from the asymmetric potential along the $c$-axis produced by the electric field (Fig. 1d), is calculated to be very small with less than 2% of the total spin polarization. This is indeed expectable by group theory, ruling that no in-plane component is allowed at the K-points owing to their three-fold rotational symmetry ($C_3$) symmetry[44, 45]. In the presence of the finite Rashba-type SOI, a Fulde–Ferrell–Larkin–Ovchinnikov (FFLO) state[4-6] (a helical state[46]), where Cooper pairs have nonzero momentum, with $s+f$-wave symmetry[47], is likely to be realized in our system. However, we confirmed by a numerical calculation that the enhancement of the upper critical field due to the FFLO state, or induced spin-triplet components derived from Zeeman-type SOI is negligible (see ref. 48 and Supplementary Fig. 6). Note that this FFLO state, where Cooper pairs have a finite center of mass momentum which is much smaller than K, should be distinguished from the intra-valley pairing. Also, as other possibilities for the enhancement of $H_{c2}$, Rashba-type SOI[1-3], quantum critical point[49] and modified electron $g$-factor[1, 30] can be ruled out in the present system (details in Supplementary Section III). Therefore, the spin-valley locking owing to intrinsic Zeeman-type SOI is considered to be the most promising origin for the enhancement of $H_{c2}^{\parallel}$.

We theoretically estimated the realistic Pauli limit of the present system by considering both the moderately large Zeeman- and the small Rashba-type SOI. For this purpose, we constructed a simpler tight-binding model reproducing the 2DES subband structure shown in Fig. 4b (see Supplementary Section IV and Supplementary Fig. 4). Assuming isotropic $s$-wave superconductivity, we then calculated the Pauli limit in this model by solving the linearized BCS gap equation using a diagrammatic technique[50] (see Supplementary Section V) based on the 2DES subband structure. Figure 4c shows the theoretical curves of the Pauli limit in this system. Considering only the Zeeman-type SOI, the Pauli limit is considerably enhanced as it is larger than 70 T at $T = 1$ K (see also Supplementary Fig. 5). This result indicates that the moderately large valley-dependent Zeeman-type spin-splitting in the vicinity of the K points (~ 13 meV) locks the spin of singlet Cooper pairing between the K and –K valleys, namely, the Cooper pairing locked by out-of-plane spin polarization to the two opposite directions, referred to as Ising pairing (Fig. 4d), enhances $H_{c2}^{\parallel}$ much higher than the $H_{P}^{BCS}$.

（改行） By contrast, once the small Rashba-type SOI is included, the enhanced Pauli limit is considerably suppressed, indicating that the symmetrical protection by spin-valley locking is weakened (Fig. 4c). This is because the in-plane polarized spin components due to the Rashba-type SOI is much more susceptible to an external in-plane magnetic field in comparison to the out-of-plane polarized spins due to intrinsic Zeeman-type SOI. The best agreement with the experimental data is obtained for a moderate Rashba-type SOI of 10% of the Zeeman-type SOI, although such Rashba-type SOI is unlikely, according to the first-principles-based band calculations as mentioned above. We discuss three possible origins (proximity to the second layer, possible small misalignment of the pulsed magnetic fields and impurity scattering effect) for this discrepancy between the theoretical results based on a single-layer tight-binding model and experimental results in Supplementary Section VI. In addition, according to our numerical calculations including the dependence on both the carrier density and $T_{c}$, the Pauli limit is predominantly controlled by both the Zeeman-type SOI and $T_{c}$, and the contribution of Rashba-type SOI is negligibly small, in the range of the carrier densities where superconductivity realizes in this system (Supplementary Fig. 7). These results demonstrate that, by the application of a strong electric field, MoS$_2$, which is believed to be a conventional superconductor in intercalated bulk forms, becomes an unconventional 2D Ising superconductor in which Cooper pairs are protected by Zeeman-type spin-valley locking, and are thereby very robust against external magnetic fields (Fig. 4d), which results in the dramatic enhancement of the Pauli limit. Our findings thus indicate that, combined with highly crystalline materials, unprecedented exotic properties of superconductivity has become accessible by the geometrical confinement owing to a strong electric field, which suggests that electric-field-induced superconductivity offers a new platform to unveil the intrinsic nature of matter.

**Methods**

Methods and any associated references are available in the online version of the paper.

**References**

1   Bauer, E. & Sigrist, M. Non-centrosymmetric superconductors: Introduction and overview. (Springer, Berlin/Heidelberg, 2012).

2   Yip, S. K. Two-dimensional superconductivity with strong spin-orbit interaction. *Phys. Rev. B* **65**, 144508 (2001).

3   Gor'kov, L. P. & Rashba, E. I. Superconducting 2D System with lifted spin degeneracy: mixed singlet-triplet state. *Phys. Rev. Lett.* **87**, 037004 (2001).

4   Frigeri, P. A., Agterberg, D. F., Koga, A. & Sigrist, M. Superconductivity without inversion symmetry: MnSi versus CePt$_3$Si. *Phys. Rev. Lett.* **92**, 097001 (2004).

5   Fulde, P & Ferrell., R. A. Superconductivity in a strong spin-exchange field. *Phys. Rev.* **135**, A550–A563 (1964).

6   Larkin, A. I. & Ovchinnikov, Y. N. Nonuniform state of superconductors, *Sov. Phys. JETP* **20**, 762–769 (1965).

7   Rashba, E. I. Properties of semiconductors with an extremum loop 1 cyclotron and combinational resonance in amagnetic field perpendicular to the plane of the loop. *Sov. Phys.-Solid State* **2**, 1109–1122 (1960).

8   Zhu, Z. Y., Cheng, Y. C. & Schwingenschlogl, U. Giant spin-orbit-induced spin splitting in two-dimensional transition-metal dichalcogenide semiconductors. *Phys. Rev. B,* **84**, 153402 (2011).

9   Xiao, D., Liu, G.-B., Feng, W., Xu, X. & Yao, W. Coupled spin and valley physics in monolayers of MoS$_2$ and other group-VI dichalcogenides. *Phys. Rev. Lett.* **108**, 196802 (2010).

10  Kormányos, A., Zólyomi, V., Drummond, N. D. & Burkard, G. Spin-orbit coupling, quantum dots, and qubits in monolayer transition metal dichalcogenides. *Phys Rev X* **4**, 011034 (2014).

11  Mak, K. F., Lee, C.,Hone, J., Shan, J. & Heinz, T. F. Atomically thin MoS$_2$: a new

directgap semiconductor. *Phys. Rev. Lett.* **105**, 136805 (2010).

12  Wang, Q. H., Kalantar-Zadeh, K., Kis, A., Coleman, J. N., Strano, M. S. Electronics and optoelectronics of two-dimensional transition metal dichalcogenides. *Nature Nanotechnol.* **7**, 699–712 (2012).

13  Wu, W *et al*. Piezoelectricity of single-atomic-layer $MoS_2$ for energy conversion and piezotronics. *Nature* **514**, 470–474 (2014).

14  Sangwan, V. K. *et al*. Gate-tunable memristive phenomena mediated by grain boundaries in single-layer $MoS_2$. *Nature Nanotechnol.* **10**, 403–406 (2015).

15  Wu. S., *et al*. Monolayer semiconductor nanocavity lasers with ultralow thresholds. *Nature* **520**, 69–72 (2015).

16  Zhang, Y. J., Oka, T., Suzuki, R., Ye, J. T. & Iwasa, Y. Electrically switchable chiral light-emitting transistor. *Science* **344**, 725–728 (2014).

17  Mak, K. F., McGill, K. L., Park, J. & McEuen, P. L. The valley Hall effect in $MoS_2$ transistors. *Science* **344**, 1489–1492 (2014).

18  Xu, X., Yao, W. Xiao, D. & Heinz, T. F. Spin and pseudospins in layered transition metal dichalcogenides. *Nature Phys.* **10**, 343–350 (2014).

19  Cui, X. *et al*. Multi-terminal transport measurements of molybdenum disulphide using van der Waals heterostructure device platform. *Nature Nanotechnol.* **10**, 534–540 (2015)

20  Ye, J. T. *et al*. Superconducting dome in a gate-tuned band insulator. *Science.* **338**, 1193–1196 (2012).

21  Yuan, H. T. *et al*. Zeeman-type spin splitting controlled by an electric field. *Nature Phys.* **9**, 563–569 (2013).

22  Kormányos, A. *et al*. Trigonal warping, the $\Gamma$ valley, and spin-orbit coupling effects. *Phys. Rev. B* **88**, 045416 (2013).

23  Coehoorn, R. *et al*. Electronic structure of $MoSe_2$, $MoS_2$, and $WSe_2$. I. Band-structure

calculations and photoelectron spectroscopy. *Phys. Rev. B* **35**, 6195–6202 (1987).

24  Molina-S´anchez, A., Sangalli, D., Hummer, K., Marini, A. & Wirtz, L. Effect of spin-orbit interaction on the optical spectra of single-layer, double-layer, and bulk $MoS_2$. *Phys. Rev. B* **88**, 045412 (2013).

25  Ge, Y. & Liu, A. Y. Phonon-mediated superconductivity in electron-doped single-layer $MoS_2$: A first-principles prediction. *Physical Review B* **87**, 241408(R) (2013).

26  Roldan, R., Cappelluti, E & Guinea, F. Interactions and superconductivity in heavily doped $MoS_2$. *Physical Review B* **88**, 054515 (2013).

27  Yuan, N. F. Q., Mak, K. F. & Law. K. T. Possible topological superconducting phases of $MoS_2$. *Phys. Rev. Lett.* **113**, 097001 (2014).

28  Rösner, M., Haas, S. & Wehling, T. O. Phase diagram of electron-doped dichalcogenides. *Phys. Rev. B* **90**, 245105 (2014).

29  Das, T. & Dolui, K. Superconducting dome in $MoS_2$ and $TiSe_2$ generated by quasiparticle-phonon coupling. *Phys. Rev. B* **91**, 094510 (2015).

30  Tinkham, M. Introduction to Superconductivity, 2nd ed. (Dover, New York, 2004).

31  Reyren, N. *et al.* Superconducting interfaces between insulating oxides. *Science* **317**, 1196–1199 (2007).

32  Kim, M., Kozuka, Y., Bell, C., Hikita, Y., & Hwang, H. Y. Intrinsic spin-orbit coupling in superconducting δ-doped $SrTiO_3$ heterostructures *Phys. Rev. B* **86**, 085121 (2012).

33  Ueno, K. *et al.* Effective thickness of two-dimensional superconductivity in a tunable triangular quantum well of $SrTiO_3$. *Phys. Rev. B* **89**, 020508(R) (2014).

34  Woolam, J. B. & Somoano, R. B. Physics and chemistry of $MoS_2$. intercalation compounds. *Mat. Sci. Eng.* **31**, 289–295 (1977).

35  Matsuda, Y. & Shimahara, H. Fulde–Ferrell–Larkin–Ovchinnikov superconductivity near the antiferromagnetic quantum critical point, *J. Phys. Soc. Jpn.* **77**, 063705 (2008).

36 Maki, K. Effect of Pauli paramagnetism on magnetic properties of high-field superconductors. *Phys. Rev.* **148**, 362–369 (1966).

37 Werthamer, N. R., Helfand, E. & Hohenberg, P. C. Temperature and purity dependence of the superconducting critical field, $H_{c2}$. III. Electron spin and spin-orbit effects. *Phys. Rev.* **147**, 295–302 (1966).

38 Klemm, R. A., Luther, A. & Beasley, M. R. Theory of upper critical-field in layered superconductors. *Physical Review B* **12**, 877–891 (1975).

39 Cuong, N. T., Otani, M. & Okada, S. Gate-induced electron-state tuning of $MoS_2$: first-principles calculations. *J. Phys.: Condens. Matter* **26** 135001 (2014).

40 Brumme, T., Calandra, M. & Mauri, F. First-principle theory of field-effect doping in transition-metal dichalcogenides: structural properties, electronic structure, Hall coefficient, and electrical conductivity. *Phys. Rev. B* **91**, 155436 (2015).

41 Liu, G.-B., Shan, W.-Y., Yao, Y., Yao, W., & Xiao, D. Three-band tight-binding model for monolayers of group-VIB transition metal dichalcogenides. *Phys. Rev. B* **88**, 085433 (2013).

42 Kormányos, A. *et al.* $\boldsymbol{k \cdot p}$ theory for two-dimensional transition metal dichalcogenide semiconductors. *2D Mater.* **2**, 022001 (2015).

43 Wu, S. *et al.* Electrical tuning of valley magnetic moment through symmetry control in bilayer $MoS_2$. *Nat. Phys.* **9**, 149–153 (2013).

44 Dresslhaus, M. S., Dresselhaus, G. & Jorio, A. Group Theory: Application to the physics of condensed matter" (Springer, Berlin Heidelberg, 2008).

45 Oguchi, T. & Shishidou, T. The surface Rashba effect: a $k \cdot p$ perturbation approach. *J. Phys.: Condens. Matter* **21,** 092001 (2009).

46 Kaur, R. P., Agterberg, D. F. & Sigrist, M. Helical vortex phase in the noncentrosymmetric $CePt_3Si$. *Phys. Rev. Lett.* **94**, 137002 (2005).

47 Frigeri, P. A. Superconductivity in crystals without an inversion center. Ph.D. thesis, (ETH Zurich, 2005).

48 Yanase, Y. & Sigrist, M. Magnetic properties in non-centrosymmetric superconductors with and without antiferromagnetic order. *J. Phys. Soc. Jpn.* **76**, 124709 (2007).

49 Tada, Y., Kawakami, N. & Fujimoto, S. Colossal enhancement of upper critical fields in noncentrosymmetric heavy fermion superconductors near quantum criticality: $CeRhSi_3$ and $CeIrSi_3$. *Phys. Rev. Lett.* **101**, 267006 (2008).

50 Nakamura, Y & Yanase. Y. Multi-orbital superconductivity in $SrTiO_3/LaAlO_3$ interface and $SrTiO_3$ surface. *J. Phys. Soc. Jpn.* **82**, 083705 (2013).

**Acknowledgements**

We thank T. Gokuden for technical support, and M. Yoshida and K. Kikutake for fruitful discussions. Y.S. and Y. Nakamura were supported by the Japan Society for the Promotion of Science (JSPS) through a research fellowship for young scientists. This work was supported by Grant-in-Aid for Specially Promoted Research (no. 25000003) from JSPS and Grant-in-Aid for Scientific Research on Innovative Areas (no. 22103004) from MEXT of Japan.

**Author contributions**

Y.S. and Y.I. conceived the idea and designed the experiments. Y.S., Y. Nakagawa and M.O. fabricated $MoS_2$-EDLT devices. Y.S. conducted cryogenic transport measurements with PPMS set up, and analysed the data. M.S.B. carried out *ab-initio*-based tight binding supercell calculations. Y. Nakamura performed numerical calculations of the upper critical field. Y.S., Y. Kasahara and Y. Kohama carried out high field measurements in the Institute for Solid State Physics. J.T.Y. took the leadership of the initial high field experiment when he was in the University of Tokyo and RIKEN. M.T., Y. Kasahara and T.N. led physical discussions. Y.S., M.S.B., T.N., Y.Y. and Y.I. wrote the manuscript.

**Additional information**

Supplementary information is available in the online version of the paper. Reprints and permissions information is available online at www.nature.com/reprints.

Correspondence and requests for materials should be addressed to Y.S. or Y.I.

**Competing financial interests**

The authors declare no competing financial interests.

**Figure Captions**

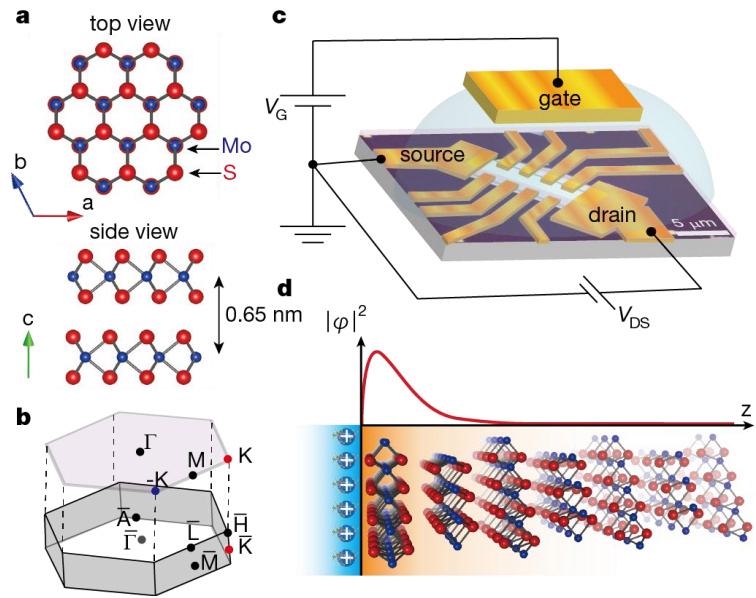

**Figure 1 | Crystal structure of MoS$_2$ and conceptual images of a MoS$_2$-EDLT. a**, Ball-and-stick model of the bulk crystal structure of MoS$_2$ for top and side view. **b**, Corresponding bulk (bottom) and monolayer (top) Brillouin zone. **c**, Schematic image of the MoS$_2$-EDLT. **d**, Schematic interface carrier profile in the MoS$_2$-EDLT.

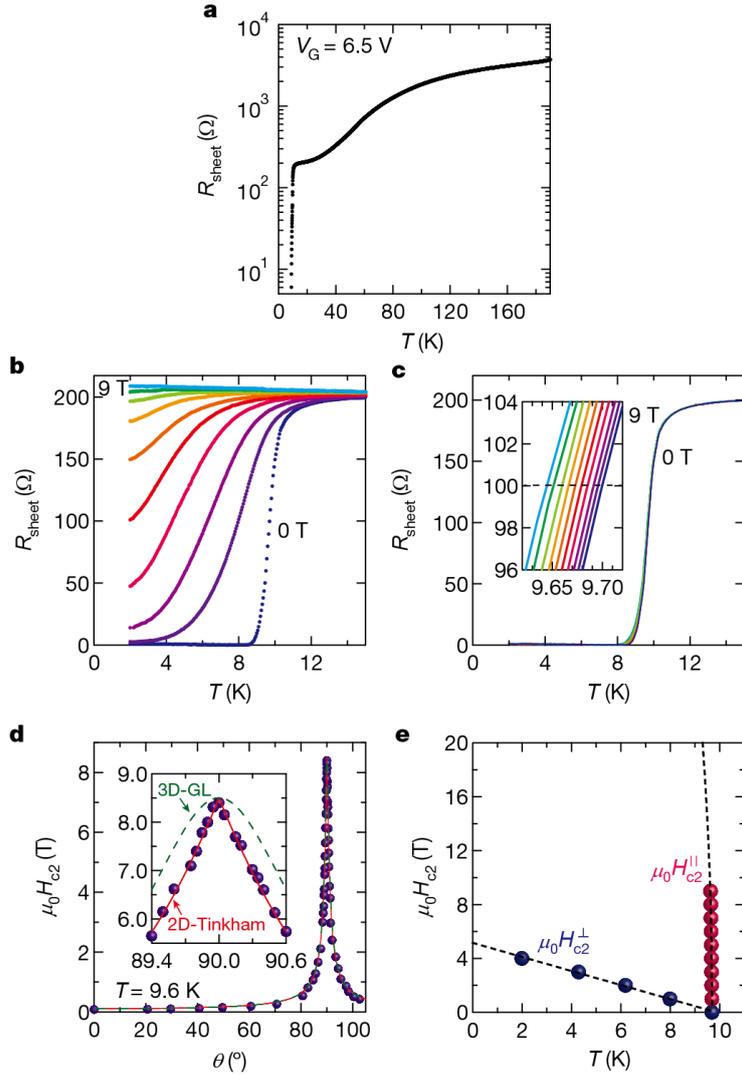

**Figure 2 | Two-dimensional superconductivity in ion-gated MoS₂. a**, Sheet resistance as a function of temperature at $V_G$ = 6.5 V. **b, c**, The superconducting transition was observed at $T$ = 9.7 K and $\mu_0 H$ = 0 T. Sheet resistance of a MoS₂-EDLT as a function of temperature at $V_G$ = 6.5 V, for (**b**) perpendicular magnetic fields, $\mu_0 H_{c2}^{\perp}$, and (**c**) parallel magnetic fields, $\mu_0 H_{c2}^{\parallel}$, varying in 1 T steps from 0 to 9 T, respectively. The inset shows a close-up of the resistive transition near the mid-point of the normal state sheet resistance (black dashed line). **d**, Angular dependence of the upper critical field, $\mu_0 H_{c2}(\theta)$ ($\theta$ is the angle between a magnetic field and the perpendicular direction to the surface of MoS₂). The inset shows a magnified view of the region around $\theta$ = 90°. The red solid line and the green dashed line correspond to the theoretical representation of $H_{c2}(\theta)$, the 2D Tinkham's formula

$\left(H_{c2}(\theta)\sin\theta \big/ H_{c2}^{\parallel}\right)^2 + \left|H_{c2}(\theta)\cos\theta \big/ H_{c2}^{\perp}\right| = 1$ and the 3D anisotropic mass model (3D-GL)

$\left(H_{c2}(\theta)\sin\theta \big/ H_{c2}^{\parallel}\right)^2 + \left(H_{c2}(\theta)\cos\theta \big/ H_{c2}^{\perp}\right)^2 = 1$, respectively. **e**, Temperature dependence of $\mu_0 H_{c2}$ perpendicular and parallel to the surface, $\mu_0 H_{c2}^{\perp}(T)$ and $\mu_0 H_{c2}^{\parallel}(T)$. Black dashed curves indicate the theoretical lines obtained from the 2D-GL equations.

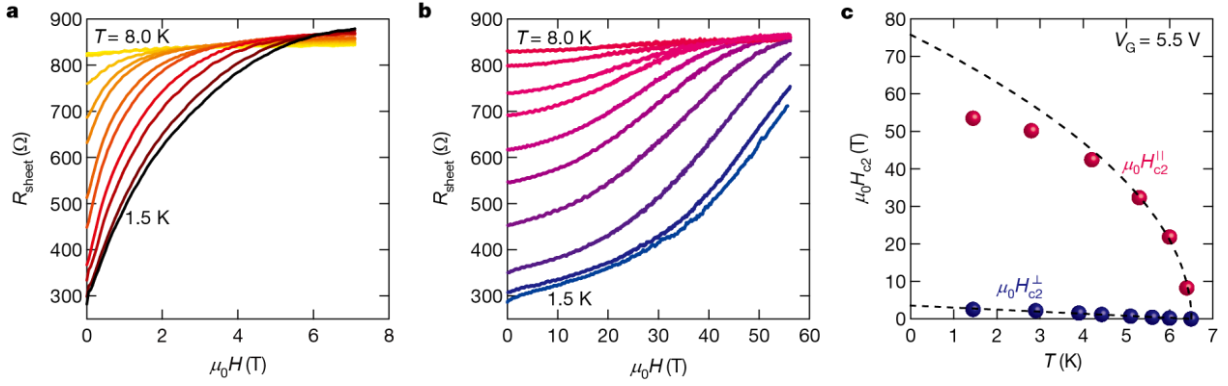

**Figure 3 | Huge upper critical fields in ion-gated MoS$_2$. a,** Sheet resistance of a MoS$_2$-EDLT as a function of magnetic field up to 55 T at $V_G$ = 5.5 V, for perpendicular magnetic fields $\mu_0H^\perp$ at 1.5, 2.9, 3.9, 4.4, 5.1, 5.6, 6, 6.5, 6.8, 7.3 and 8.0 K, and **b,** for parallel magnetic fields $\mu_0H^\parallel$ at 1.5, 2.8, 4.2, 5.3, 6, 6.4, 6.5, 6.8, 7.1, 7.6 and 8.0 K. **c,** In-plane and out-of-plane upper critical fields as a function of temperature. $H_{c2}$ is defined by the magnetic field where $R_{sheet}$ reached 75% of the normal state sheet resistance. The black dashed curves show the 2D-GL model. The $H_{c2}^\parallel$ increases with decreasing temperature followed by the 2D-GL model near $T_c$, but it deviates from the model at lower temperatures and eventually saturates approximately 52 T at 1.5 K, suggestive of the enhancement of the Pauli limit.

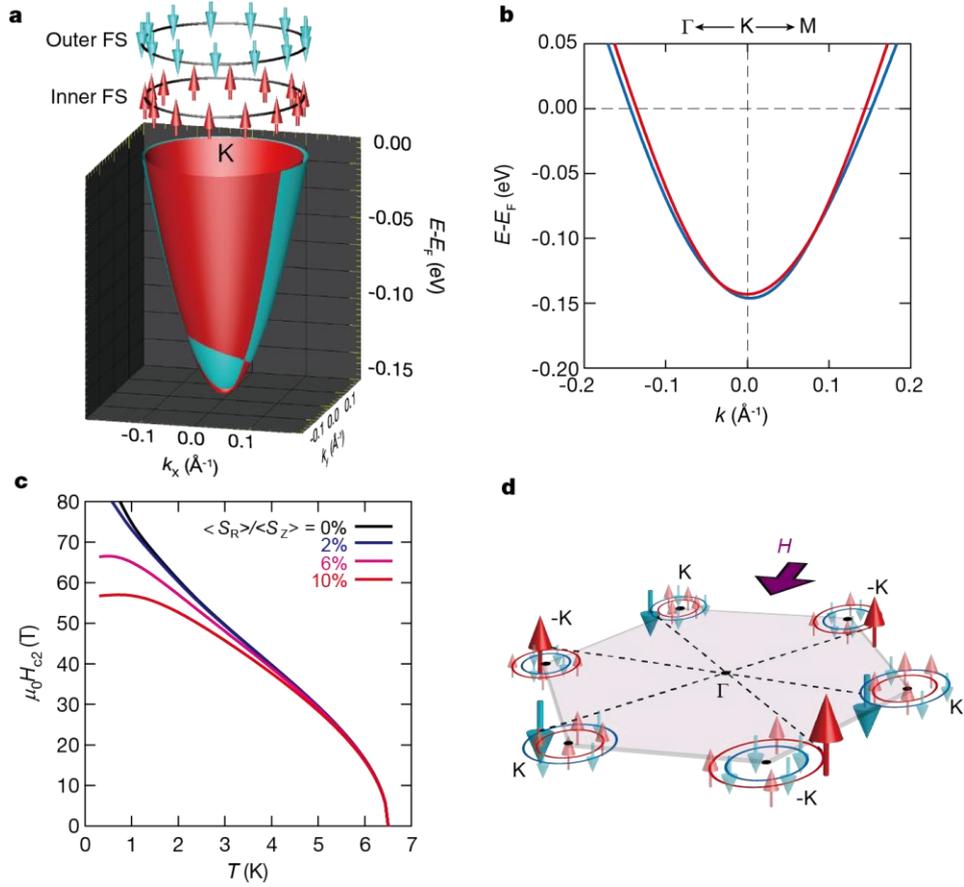

**Figure 4 | Ising paring protected by spin-valley locking in electric-field-induced 2D Ising superconductivity in MoS₂. a**, Energy band dispersion and spin texture of the conduction band around the K point of bulk MoS₂ under a strong electric field at $n_{2D} = 8.7 \times 10^{13}$ cm⁻². Inner Fermi surface (FS) and outer FS at the K points have out-of-plane spin-polarization with opposite directions because each band is almost fully out-of-plane spin polarized by effective valley Zeeman fields, while the in-plane Rashba-type component is very small with less than 2% of the total spin polarization. **b**, Two-dimensional energy band dispersion near the K point. The spin-split bands cross with each other. The splitting at the Fermi level becomes ~ 13 meV, while that at the K point is ~ 3 meV. **c**, Theoretical curves of the Pauli limit considering both the moderately large Zeeman- and small Rashba-type SOI based on 2DES subband structure (see also Supplementary Figs. 5 and 7). Black curve is the upper critical field in the tight-binding model reproducing the band structure calculation (Supplementary Fig. 4). The ratio of the Rashba- and Zeeman-type SOI, $\langle S_R \rangle / \langle S_Z \rangle$, is varied

from 0% to 10%. **d**, Schematic image of the Fermi surfaces with valley-dependent spin-polarization in the in-plane magnetic field geometry. The direction of each spin is orthogonal to the magnetic field. Inter-valley Ising pairing formed between the K and –K valleys is robust against an external magnetic field $H$, which realizes spin-valley coupled 2D Ising superconductivity in ion-gated $MoS_2$.

**Methods**

**Device fabrication.** Bulk 2$H$-polytype $MoS_2$ single crystals were cleaved into thin flakes with dozens of nanometers in thickness using the Scotch-tape method, and then, flakes were transferred onto a $Si/SiO_2$ substrate or a Nb-doped $SrTiO_3/HfO_2$ substrate. Au (90 nm)/Cr (5 nm) electrodes were patterned onto an isolated thin flake in a Hall bar configuration, and a side gate electrode was patterned onto the substrate. We covered the device with ZEP 520A (used as the resist for electron beam lithography) except for the channel surface, to avoid chemical intercalation from the edge of the flake, allowing us to focus on the field effect. A droplet of ionic liquid covered both the channel area and the gate electrode. Ionic liquid N,N-diethyl-N-(2-methoxyethyl)-N- methylammonium bis (trifluoromethylsulphonyl) imide (DEME-TFSI) was selected as a gate medium.

**Transport measurements.** The temperature dependent resistance under magnetic fields, of the $MoS_2$-EDLT (shown in Fig. 2) was measured with a standard four-probe geometry in a Quantum Design Physical Property Measurement System (PPMS) with the Horizontal Rotator Probe with an error below 0.01°, combined with two kinds of AC lock-in amplifiers (Stanford Research Systems Model SR830 DSP lock-in amplifier and Signal Recovery Model 5210 lock-in amplifier). The gate voltage was supplied by a Keithley 2400 source meter. We applied gate voltages to the device at 220 K, which is just above the glass transition temperature of DEME-TFSI, under high vacuum (less than $10^{-4}$ Torr), and cooled down to low temperatures. The excitation source-drain current used in the PPMS set up was limited to 1 $\mu$A to avoid heating and large-current effect to superconductivity.

# Supplementary Information for

# Superconductivity protected by spin-valley locking in ion-gated MoS$_2$

Yu Saito*, Yasuharu Nakamura, Mohammad Saeed Bahramy, Yoshimitsu Kohama,

Jianting Ye, Yuichi Kasahara, Yuji Nakagawa, Masaru Onga, Masashi Tokunaga,

Tsutomu Nojima, Youichi Yanase and Yoshihiro Iwasa*

*Corresponding author. E-mail: saito@mp.t.u-tokyo.ac.jp (Y.S.); iwasa@ap.t.u-tokyo.ac.jp (Y.I.)

## Contents

# I. Possibility of electrochemical reaction

In the entire measurement, the electrochemical reaction can be excluded for the following reasons.

First of all, at low temperatures, especially just above the glass transition temperature of ionic liquid, which is approximately 190 K, the activation energy of electrochemical process is significantly suppressed. Supplementary Figure 1a displays the temperature dependent gate current, $I_{Gate}$, in a MoS$_2$-EDLT. $I_{Gate}$ shows an activation-type reduction with temperature, with an activation energy of ~ 0.4 eV. This dramatic reduction of $I_{Gate}$ upon lowering temperature suggests that the electrochemical window is effectively widened[1]. Also, it has been reported that the maximum gate voltage applicable increases from 3 to 5.5 V, simply by decreasing temperature from 300 to 220 K[2].

Furthermore, to confirm the electrostatic process of ionic-liquid gating, we measured transfer curves of a MoS$_2$-EDLT at 220 K. Supplementary Figure 1b shows the source-drain current, $I_{DS}$, and $I_{Gate}$ as a function of gate voltage between -1 and 6.5 V. Both $I_{DS}$ and $I_{Gate}$ of the device completely returned back to the original values without large hysteresis. This is a strong indication that the ionic-liquid gating is a reversible process, and thus the electrochemical reaction is highly unlikely.

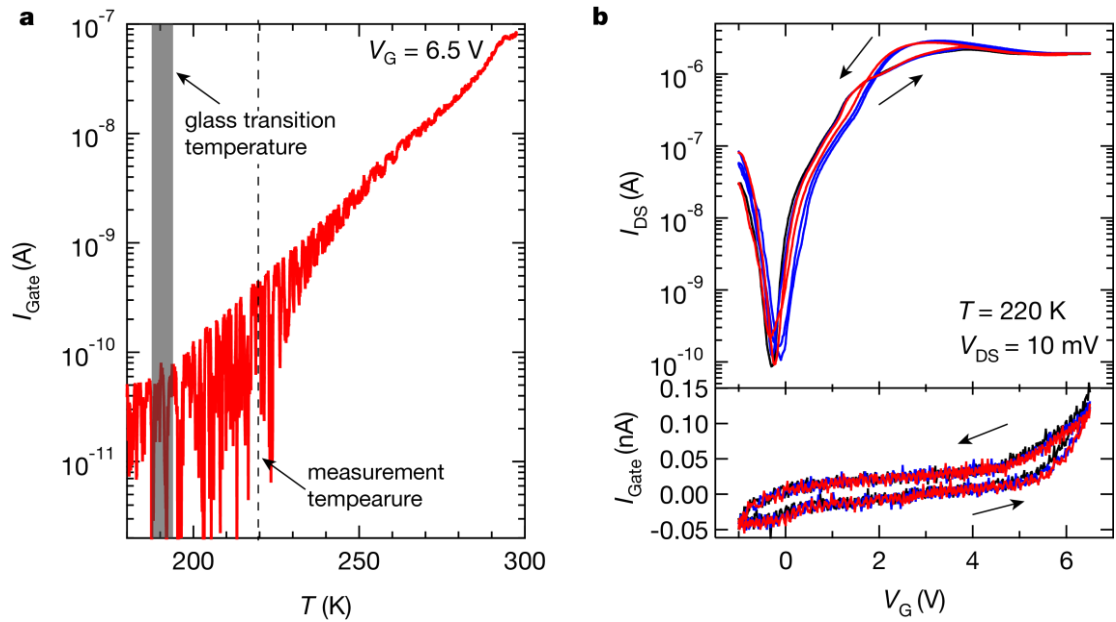

**Supplementary Figure 1 | Temperature dependence of (a) the gate current $I_{Gate}$ and (b) the transfer curve at 220 K of a MoS$_2$-EDLT.**

## II. High magnetic field measurements

We performed all the magnetoresistance measurements in a four-probe configuration using a 55 T pulsed magnet with the duration time of 36 ms and a rotator probe at the International MegaGauss Science Laboratory, Institute for Solid State Physics, the University of Tokyo. The voltage signals were recorded on National Instruments PXIe-6124 digitizers using a homemade numerical lock-in technique at a frequency of $f = 80$ kHz. The sensitivity of the angle measurement is within ~ 0.1 degree in our rotator probe. To align the field direction, we performed the $H_{c2}$ measurement as a function of angle in pulsed magnetic fields. The misalignments for the in-plane magnetic fields and out-of-plane magnetic fields are estimated to be less than ~ 0.02 and ~ 1.0 degree, respectively. We observed clear voltage signals of the magnetoresistance from a $MoS_2$-EDLT, followed by pulsed magnetic fields. Supplementary Figure 2 shows a typical time dependent voltage signal under a pulsed magnetic field. The magnetoresistance data (shown in Fig. 3, a and b) was obtained during the down sweep of a field pulse with the field perpendicular to the $c$-axis. All these magnetoresistance measurements have been performed out under the condition that the source-drain current, $I_{DS}$, was less than 10 $\mu$A, in which we confirmed that the behaviour of the temperature-dependent resistance and magnetoresistance below 8 K were almost unchanged. Some eddy current heating was observed in measurements between 4 and 8 K. Below 4 K, the $MoS_2$-EDLT was immersed in liquid helium, which prevents heating in the device.

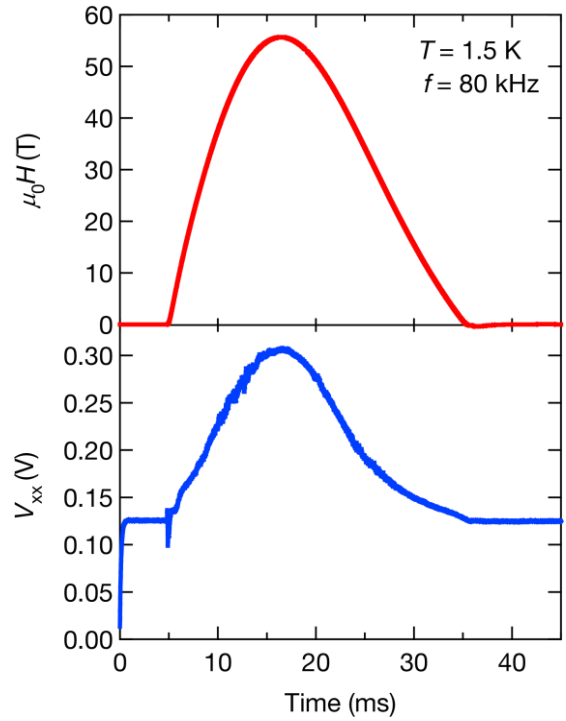

**Supplementary Figure 2 | Time-dependent magnetic field and voltage probe signal in a MoS$_2$-EDLT at $V_G$ = 5.5 V.** The upper and lower panels show a pulsed magnetic field and the voltage probe signal as a function of time, respectively. This measurement was made at $T$ = 1.5 K with an AC measurement set up with a frequency of $f$ = 80 kHz.

## III. Possible origins of the enhancement of $H_{c2}$

### 1) Spin orbit scattering

To discuss spin orbit scattering effect on the enhancement of $H_{c2}$ on MoS$_2$-EDLTs, we fitted $H_{c2}^{\parallel}(T)$ values of three devices (Device H1, H2, L1; those devices showed different $T_c$ at different carrier densities (Supplementary Fig. 3a).) by using the microscopic KLB theory[4]. In this theory, $H_{c2}^{\parallel}(T)$ arising from a monolayer satisfies the following equations,

$$\ln\left(\frac{T}{T_c}\right) + \psi\left(\frac{1}{2} + \frac{3\tau_{SO}\left(\mu_B H_{c2}^{\parallel}\right)^2}{4\pi T}\right) - \psi\left(\frac{1}{2}\right) = 0$$

where $\psi(x)$, $\mu_B$ and $T_c$ are digamma function, the Bohr magnetism and $T_c$, respectively. Using this equation, we fitted the experimental data (Supplementary Fig. 3, b and c) and then estimated the values of $\tau_{SO}$ as a fitting parameter (Supplementary Table 1). Although all $H_{c2}^{\parallel}(T)$ values seem to be quite well fitted by the KLB theory, $\tau$ estimated by the transport is larger than $\tau_{SO}$, being unphysical. Thus, spin orbit scattering is unlikely to be responsible for the enhancement of $H_{c2}^{\parallel}(T)$.

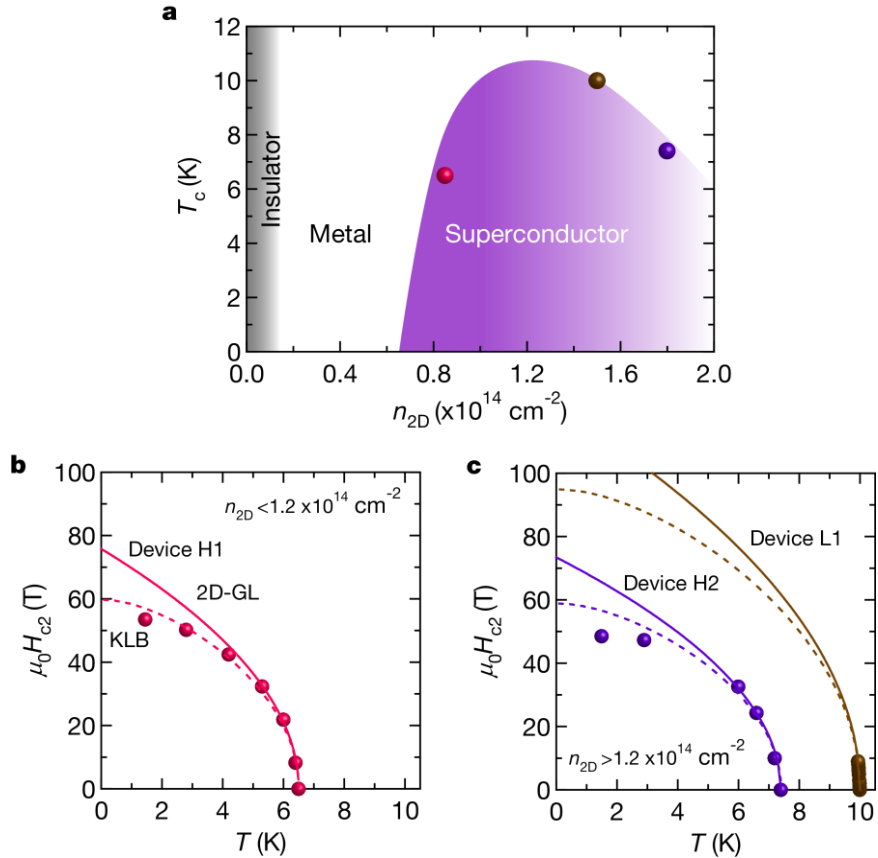

**Supplementary Figure 3 |** $H_{c2}^{\parallel}(T)$ **values at different carrier densities and fitting results.**
**a**. Electronic phase diagram of ion-gated MoS$_2$. Here, $T_c$ is determined at the temperature where the resistance drop reaches 75% of its normal state sheet resistance $R_N$ at 15 K. **b, c,** Temperature dependence of $H_{c2}^{\parallel}(T)$ at different carrier densities (**b**: $n_{2D} < 1.2 \times 10^{14}$ cm$^{-2}$, **c**: $n_{2D} > 1.2 \times 10^{14}$ cm$^{-2}$). Solid and dashed curves show the 2D-GL and the KLB fittings, respectively.

|  | $T_c$ (K) | $n_{2D}$ (cm$^{-2}$) | $\mu_H$ (cm$^2$/Vs) | $\tau$ (fs) | $\tau_{SO}$ (fs) |
|---|---|---|---|---|---|
| Device L1 | 10.0 | $1.5 \times 10^{14}$ | 208 | 59.3 | 11.1 |
| Device H1 | 6.5 | $0.85 \times 10^{14}$ | 86 | 25.5 | 17.9 |
| Device H2 | 7.4 | $1.8 \times 10^{14}$ | 165 | 47.1 | 21.3 |

**Supplementary Table 1 | Device properties of MoS$_2$-EDLTs.** $n_{2D}$, Hall mobility, $\mu_H$, and $\tau$ are values measured at 15 K.

## 2) Rashba SOI

Rashba spin-orbit coupling (SOC) will lock the spin to the in-plane direction, which can enhance the out-of-plane $H_{c2}$ beyond the Pauli limit[5]. However, for in-plane magnetic fields, most of the electron spins can contribute to Pauli paramagnetism. Therefore, the enhancement of $H_{c2}$ can be only $\sqrt{2}H_P^{BCS}$[6]. In the present case, $H_{c2}$ is enhanced by a factor of 4, which is much larger than the enhancement due to the Rashba SOI. Furthermore, according to our calculations, the Rashba SOI is negligibly small in the present system, and thus the pure Rashba SOI effect does not contribute to the enhancement of $H_{c2}$

## 3) Quantum critical point

No ordered state (ex. antiferromagnetic state) has been observed in ion-gated MoS$_2$ system in the vicinity of the superconducting phase. In that narrow sense, ion-gated MoS$_2$ system has no quantum critical point (QCP), which is known to dramatically enhance the upper critical field[7]. Thus, the enhancement of $H_{c2}$ by the QCP can be ruled out.

## 4) Modified electron *g*-factor

The enhancement of upper critical fields owing to modified electron g-factor may become effective in the case that the LS coupling of multiple *d*-orbitals competing with crystal field splitting[5, 8] stabilizes the spin-orbital coupled ground state. In the present system, the conduction band at the Fermi level is composed of only single *d* orbital $(d_{z^2})^9$, and therefore, the effect of modified electron *g*-factor is supposed to be negligible.

## IV. Tight-binding model

To calculate the subband structure of $MoS_2$ under a strong electric field, we first carried out density functional theory calculations for bulk $MoS_2$ using the full potential augmented plane-wave method and Perdew–Burke–Ernzerhof exchange-correlation functional modified by Becke–Johnson potential, as implemented in WIEN2K program[10]. For each atom, the muffin-tin radius, $R_{MT}$, was chosen such that its product with the maximum modulus of reciprocal vectors, $K_{max}$, becomes $R_{MT}\,K_{max} = 7.0$. The relativistic effects, including spin-orbit interaction were fully included and the Brillouin zone was sampled by a $12 \times 12 \times 6$ $k$-mesh. We then downfolded the bulk Hamiltonian using maximally localized Wannier functions[11-13] and generated a large tight-binding supercell Hamiltonian with an additional potential term to account for the band bending. Finally, we solved this self-consistently using the Poisson equation.

The calculation of $H_{c2}^{||}$ was carried out based on the tight-binding model which reproduces the conduction band of the effectively single layer $MoS_2$. The single particle component of the Hamiltonian is given by $H = H_{kin} + H_Z + H_R$, where the first term is the kinetic energy term, $H_{kin} = \sum_{ks} \varepsilon(\mathbf{k}) c_{ks}^\dagger c_{ks}$. The dispersion relation obtained by the first-principles-based band structure calculation (shown in Fig. 4, a and b) is reproduced well by taking into account the nearest-, next-nearest-, and third-nearest-neighbour hopping,

$$
\begin{aligned}
\varepsilon(\mathbf{k}) = {} & 2t_1\left(\cos k_y a + 2\cos\frac{\sqrt{3}}{2}k_x a \cos\frac{1}{2}k_y a\right) \\
& + 2t_2\left(\cos\sqrt{3}k_x a + 2\cos\frac{\sqrt{3}}{2}k_x a \cos\frac{3}{2}k_y a\right) \\
& + 2t_3\left(\cos 2k_y a + 2\cos\sqrt{3}k_x a \cos k_y a\right) - \mu.
\end{aligned}
$$

where $a$ is the lattice constant. We determined the hopping parameters $t_1$, $t_2$, and $t_3$, and the chemical potential $\mu$ so as to reproduce the band structure calculation. The Zeeman-type SOI arising from the intrinsic inversion symmetry breaking in the crystal structure of $MoS_2$ is represented by $H_z$, whereas the Rashba-type SOI induced by the extrinsic electric field is represented by $H_R$. They are characterized by the g-vector as $H_Z = \alpha_Z \sum_{kss'} \mathbf{g}_Z(\mathbf{k})\boldsymbol{\sigma}_{ss'} c_{ks}^\dagger c_{ks'}$ and $H_R = \alpha_R \sum_{kss'} \mathbf{g}_R(\mathbf{k})\boldsymbol{\sigma}_{ss'} c_{ks}^\dagger c_{ks'}$, where

$$\mathbf{g}_Z(\mathbf{k}) = F(\mathbf{k})\left(0, 0, \sin k_y a - 2\cos\frac{\sqrt{3}}{2}k_x a \sin\frac{1}{2}k_y a\right), \qquad \text{and}$$

$$\mathbf{g}_R(\mathbf{k}) = F(\mathbf{k})\left(-\sin k_y a - \cos\frac{\sqrt{3}}{2}k_x a \sin\frac{1}{2}k_y a, \sqrt{3}\sin\frac{\sqrt{3}}{2}k_x a \cos\frac{1}{2}k_y a, 0\right).$$

We introduced a trial function $F(\mathbf{k}) = \beta \tanh\big[f(\mathbf{K}) - f(\mathbf{k})\big] - 1$ in order to reproduce the sign change of the spin-splitting from the K point to the $\Gamma$ point in the Brillouin zone. $\mathbf{K} = \left(0, \frac{4\pi}{3a}\right)$ is the wave vector at the K point. We chose the symmetric function $f(\mathbf{k}) = \left|\sin k_y a - 2\cos\frac{\sqrt{3}}{2}k_x a \sin\frac{1}{2}k_y a\right|$ for simplicity. By diagonalizing the single particle Hamiltonian, we obtained the spin-split bands $E_+(\mathbf{k})$ and $E_-(\mathbf{k})$, and fitted parameters $\alpha_Z, \alpha_R$, and $\beta$ to the results of band structure calculation, $E_+(\mathbf{k}_F) - E_-(\mathbf{k}_F) = 13$ meV, $E_+(\mathbf{K}) - E_-(\mathbf{K}) = 3$ meV at $n_{2D} = 8.7 \times 10^{13}$ cm$^{-2}$, and the ratio of the Rashba- and Zeeman-type SOIs $\langle S_R \rangle / \langle S_Z \rangle = |\alpha_R \mathbf{g}_R(\mathbf{k}_F)| / |\alpha_Z \mathbf{g}_Z(\mathbf{k}_F)| = 0.02$. The last equation relies on the results of *ab-intio*-based band calculations that the Rashba-type spin-splitting is at most 2% of the Zeeman-type spin-splitting on the Fermi surface. We defined the Fermi momentum along the K-M line, $\mathbf{k}_F$, where $E_+(\mathbf{k}_F) + E_-(\mathbf{k}_F) = 0$. Adopting these parameters, we obtained the band structure in Supplementary Fig. 4, which matches with Fig. 4b obtained by the band structure calculation for the MoS$_2$-EDLT. In order to clarify the roles of Rashba-type SOI, we varied $\langle S_R \rangle / \langle S_Z \rangle$ from 0% to 10% in Fig. 4c.

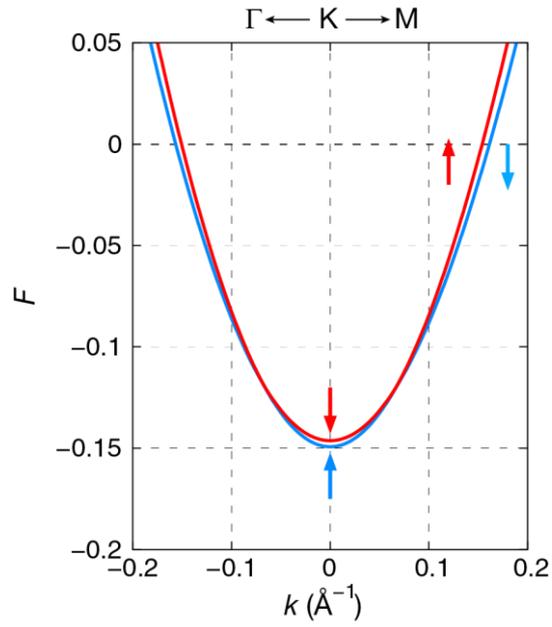

**Supplementary Figure 4 | Band structure near the K point reproduced by the tight-binding model.** The horizontal axis shows the deviation from the K point, $\pm|\mathbf{k}-\mathbf{K}|$. The positive sign + is adopted for the momentum on the K-M line, whereas the negative sign - is chosen for the momentum on the K-$\Gamma$ line.

## V. Numerical calculation of the Pauli limit

We calculated the Pauli limit, in the present system, based on the BCS model $H = H_0 + H_P + H_I$. Taking into account the Zeeman field by an external magnetic field, $H_P = -\sum_{\mathbf{k}ss'} \mu_B \mathbf{H} \cdot \boldsymbol{\sigma}_{ss'} c_{\mathbf{k}s}^\dagger c_{\mathbf{k}s'}$, and the pairing interaction in the $s$-wave channel $H_I = V \sum_i n_{i\uparrow} n_{i\downarrow}$, we solved the linearized gap equation $1 - V\chi_{sc}^0(T_c, H_{c2}) = 0$ where $V$ is the pairing interaction and the irreducible superconducting susceptibility as a function of temperature and magnetic field, and determine both $T_c$ and $H_{c2}$. We focused on the in-plane magnetic field $\mathbf{H} = H\hat{x}$. The irreducible superconducting susceptibility is obtained as

$$\chi_{sc}^0(T,H) = T\sum_{\omega_l}\sum_{\mathbf{k}}\Big[ G_{\uparrow\uparrow}(\mathbf{k}, i\omega_l)G_{\downarrow\downarrow}(-\mathbf{k}, -i\omega_l) - G_{\downarrow\uparrow}(\mathbf{k}, i\omega_l)G_{\downarrow\uparrow}(-\mathbf{k}, -i\omega_l) \Big]$$

by using the Matsubara Green function, $\hat{G}(k) = \left( i\omega_l - \hat{H}_0 - \hat{H}_P \right)^{-1}$. We assumed an attractive interaction $V/t_1 = -2.32$ so that the transition temperature of $T_c = 6.5$ K experimentally observed at zero magnetic field was reproduced.

Supplementary Figure 5 shows the theoretical curves of the Pauli limit in this system. It is shown that the Pauli limit is enhanced by the Zeeman-type SOI to over 70 T at $T = 1$ K. The Zeeman-type spin splitting at the Fermi level of ~ 13 meV plays an essential role in the Pauli limit much higher than the BCS value. Indeed, the smaller Pauli limit than the experimental values is obtained, when we assume a Zeeman-type spin splitting of 3 meV.

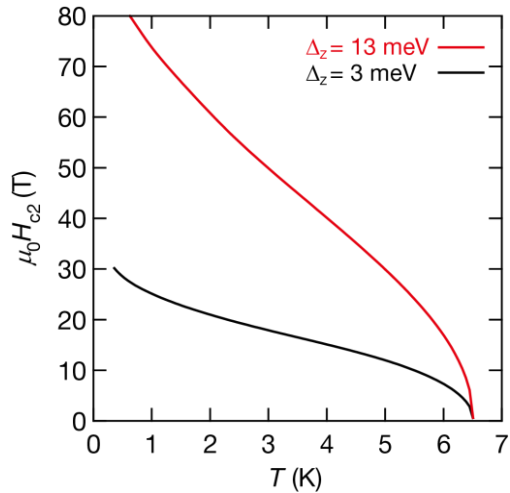

**Supplementary Figure 5 | Theoretical curves of the Pauli limit**. Red and black curves show the calculated upper critical fields assuming Zeeman-type splittings, $\Delta_Z$, of 13 and 3 meV, respectively.

Our theoretical calculation indeed shows that a FFLO superconducting state (or a helical state) in inter-valley pairing is stabilized in the presence of the Rashba-type SOI. To check the enhancement by the FFLO state, we have performed the numerical calculation of the Pauli limit at both BCS state and FFLO state (Supplementary Fig. 6). Our numerical calculation suggests that the difference between two cases is indeed negligible, and thus the enhancement of the upper critical field due to the FFLO state in our case is negligible by the numerical calculation as shown below, because the Rashba component is extremely small.

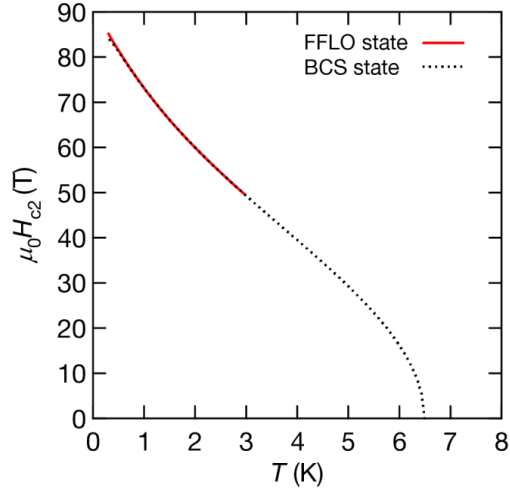

**Supplementary Figure 6 | Theoretical values of the Pauli limit at the BCS state and a FFLO state.**

To discuss the carrier density dependence of $H_{c2}$ in more detail, we show the theoretically calculated value of the Pauli limit as a function of temperature at different carrier densities and $T_c$ as shown in Supplementary Fig. 7. Here, the Rashba component is fixed at 2% of the total spin polarization, as we find no meaningful change in the behaviour of the Pauli limit for Rashba components less than this value. To calculate the Pauli limit, we have also used the same values of the spin splitting which were obtained by our *ab-intio*-based tight-binding supercell calculations and estimated from those values. The spin splitting approximately varies from 9 to 15 eV in the range of the carrier density where superconductivity realizes ($n_{2D} \sim 0.6 - 1.8 \times 10^{14}$ cm$^{-2}$), and, in this regime, the K points are much dominantly occupied. As Supplementary Fig. 7 illustrates, the temperature dependence of the Pauli limit at $n_{2D} = 8.7 \times 10^{13}$ cm$^{-2}$ is similar to that at $1.8 \times 10^{14}$ cm$^{-2}$, although their carrier densities are much different. This similarity is in good agreement with the experimental result (Supplementary Fig. 3, a and b). Combined with experimental data, we concluded that the behaviour of the upper critical field is predominantly controlled by the Zeeman-type SOI and $T_c$ in this range of the carrier density.

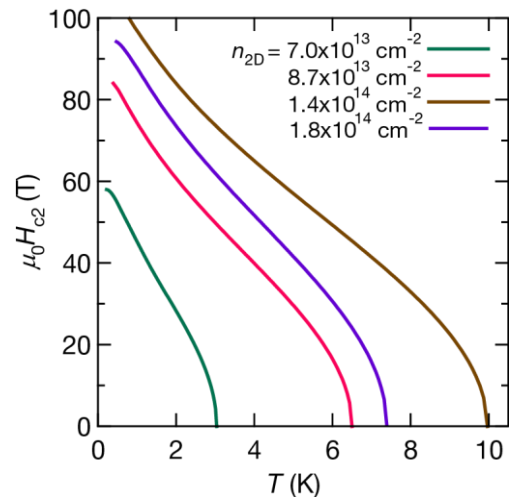

**Supplementary Figure 7 | Theoretical curves of the Pauli limit at different carrier densities and $T_c$.**

## VI. Discrepancy between the theoretical results and experimental results

We here discuss three possible origins for the discrepancy between the theoretical results based on a single-layer tight-binding model and experimental results. One is the weak proximity of carriers to the second $MoS_2$ layer, in which small amount of carriers is considered to be accumulated. Because the second layer acquires Zeeman-type SOI with an opposite sign to the first layer, the proximity leads to a suppression of SOI and suppresses the enhancement of the Pauli limit. The second possibility is a small misalignment of the pulsed magnetic fields. The pulsed magnetic fields may not be precisely parallel to the surface of the channel, because of an error in the angle or the non-uniformity of pulsed magnetic fields. However, taking the size of device into account, the non-uniformity of pulsed magnetic fields is considered to be less than 0.1 T. The third possibility is impurity scattering effect, which is known to suppress the upper critical field[14]. This effect is not included in our calculation for simplicity.

## References

1. Yuan, H. T. *et al.* High-density carrier accumulation in ZnO field-effect transistors gated by electric double layers of ionic liquids. *Adv. Funct. Mater.* **19**, 1046–1053 (2009).

2. Yuan, H. T. *et al.* Electrostatic and electrochemical nature of liquid-gated electric-double-layer transistors based on oxide semiconductors. *J. Am. Chem. Soc.* **132**, 18402 (2010).

3. Ye, J. T. *et al.* Superconducting dome in a gate-tuned band insulator. *Science.* **338**, 1193–1196 (2012).

4. Klemm, R. A., Luther, A. & Beasley, M. R. Theory of upper critical-field in layered superconductors. *Physical Review B* **12**, 877–891 (1975).

5. Bauer, E. & Sigrist, M. Non-centrosymmetric superconductors: Introduction and overview. (Springer, Berlin/Heidelberg, 2012).

6. Gor'kov, L. P. & Rashba, E. I. Superconducting 2D system with lifted spin degeneracy: mixed singlet-triplet state. *Phys. Rev. Lett.* **87**, 037004 (2001).

7. Tada, Y., Kawakami, N. & Fujimoto, S. Colossal enhancement of upper critical fields in noncentrosymmetric heavy fermion superconductors near quantum criticality: $CeRhSi_3$ and $CeIrSi_3$. *Phys. Rev. Lett.* **101**, 267006 (2008).

8. Tinkham, M. Introduction to Superconductivity, 2nd ed. (Dover, New York, 2004).

9. Zhu, Z. Y., Cheng, Y. C. & Schwingenschlogl, U. Giant spin-orbit-induced spin splitting in two-dimensional transition-metal dichalcogenide semiconductors. *Phys. Rev. B,* **84**, 153402 (2011).

10. Blaha, P., Schwarz, K., Madsen, G., Kvanicka, D. & J. Luitz. WIEN2K program package. Available at http://www.wien2k.at.

11. Souza, I. *et al.* Maximally localized Wannier functions for entangled energy bands. *Phys. Rev. B* **65**, 035109 (2001).

12. Mostofi, A. A. *et al.* Wannier90: a tool for obtaining maximally localized Wannier functions. *Comp. Phys. Commun.* **178**, 685–699 (2008).

13. Kunes, J. *et al.* WIEN2WANNIER: from linearized augmented plane waves to maximally localized Wannier functions. *Comp. Phys. Commun.* **181**, 1888–1895 (2010).

14. Bulaevskii, L. N., Guselnov, A. A. & Rusinov, A. I. Superconductivity in crystals without symmetry centers. *Sov. Phys. JETP* **44**, 1243–1251 (1976).